\documentstyle[aps,preprint]{revtex}
\newcommand{\journal}[4]{{{\sl #1}} {\bf #2}, {#3} (#4)}
\newcommand{\mprb}[3]{\journal{Phys.~Rev.~B}{#1}{#2}{#3}}
\newcommand{\mpra}[3]{\journal{Phys.~Rev.~A}{#1}{#2}{#3}}
\newcommand{\mprl}[3]{\journal{Phys.~Rev.~Lett.~}{#1}{#2}{#3}}

\def\rvec{\mbox{\boldmath $r$}}
\def\qvec{\mbox{\boldmath $q$}}

\def\pa{\partial}
 
\def\o{\omega} 
\def\d{\delta}
\def\cf{{\cal F}}

\begin{document}
\title{
Dynamics of Ordering in Alloys\\
with Modulated Phases}
\draft
\author{Bulbul Chakraborty}
\address{
The Martin Fisher School of Physics\\
Brandeis University\\
Waltham, MA 02254, USA}
\date{\today}
\maketitle

\begin{abstract}
This paper presents a theoretical model for studying the dynamics of
ordering in
alloys which exhibit modulated phases.  The model is different from the
standard
time-dependent Ginzburg-Landau description of the evolution of a
non-conserved order
parameter and resembles the Swift-Hohenberg  model.
The early-stage growth kinetics is analyzed and compared to the
Cahn-Hilliard
theory of continuous ordering.  The effects of non-linearities on the
growth
kinetics are discussed qualitatively and  it is shown that the presence
of an
underlying elastic lattice introduces qualitatively new effects.  A lattice
Hamiltonian capable of describing these effects  and suitable for
carrying out
simulations of the growth kinetics is also constructed.
\end{abstract}
\narrowtext

\section{Introduction}

One of the most fascinating problems in statistical mechanics is that of
growth kinetics \cite{Gunton}; the evolution of order following a quench
from an initial high temperature state to one below the
transition temperature.  Theoretical studies of kinetics have mainly
concentrated on understanding the behavior of models such as the
time-dependent Ginzburg--Landau
model \cite{Mazenko} or the kinetic Ising model \cite{mcising}.
Binary alloys have been viewed as convenient
model systems for experimental investigation of these theoretical
predictions \cite{Nagler,Ludwig}.
Many alloys, however,
exhibit complex ordered phases including long-period, modulated
structures \cite{modul} which cannot be described
by simple
antiferromagnetic Ising models \cite{selke}.
The kinetics of ordering in these alloys should be significantly different
from
the usual models of ordering.  These alloys can in turn provide models
for
both theoretical and experimental studies of a different class of kinetic
phenomena.

This paper presents model partial-differential
equations for describing growth kinetics in alloys with modulated
phases and analyzes the early stage kinetics.  The kinetics of growth
in the strongly non-linear regime have to be approached through
solutions of the partial differential equations or through Monte
Carlo simulation of a discrete lattice Hamiltonian.  Such a discrete
Hamiltonian, derived from the same atomistic description that leads to
the partial differential equations, is also presented, and the
qualitative aspects of
this model are discussed.

In recent years, there has been a concerted effort to understand phase
stability in metallic alloys based on a realistic description of the
``electron glue" responsible for metallic cohesion.  One such approach,
which has recently been applied to equilibrium phase transitions in the
Cu-Au
alloys with success \cite{prl,Xi},  is based
on the effective medium theory (EMT) of cohesion in metals \cite{emt}.
EMT provides a semi-empirical scheme for writing down the energy of
a system of electrons and ions with an {\it arbitrary} configuration.
This is its major advantage over first-principles approaches which
have to rely on specific symmetries of the system such as ~a) completely
random \cite{Gyorffy} or ~b) ordered periodic structures \cite{Zunger}.
By its very nature, EMT is more approximate than the first-principles
approaches; however, it does provide a model Hamiltonian
which takes into account the metallic character of these alloys,
describes the coupling of positional and configuration degrees of
freedom \cite{prl,Xi} and can deal with size effects arising from the fact
that the
Ising spins are actually atoms which have different ``sizes".

The EMT expression for the total energy for a given configuration of
ions
provides the classical Hamiltonian describing the alloys \cite{prl,Xi}.
This Hamiltonian is a function of both the positions of the atoms and
their chemical identity.  The latter can be
represented by an Ising variable \cite{revalloy}.  So far, our studies of
alloys have been based on an EMT Hamiltonian which includes the
Ising
variables and uniform lattice distortions, {\it ie} volume and shape
distortions have been included but not volume conserving
phonons \cite{prl,Xi}.  From studies of compressible Ising systems, it
is known that the coupling to uniform lattice distortions can change the
nature of the Ising phase transition \cite{halperin,MC}.  The most
striking demonstration of this is an Ising antiferromagnet on a
deformable triangular lattice where the system was found to order,
through a first-order phase transition, into a
striped phase accompanied by a lattice distortion which has the
antiferromagnetic bonds shortened and the ferromagnetic bonds
lengthened \cite{Kardar}.
On a rigid lattice there is no
ordering.  Our studies of the ordering in $CuAu$, based on EMT with
uniform lattice distortions, showed that including the lattice
distortions leads to a first-order transition from the disordered,
undistorted
lattice to a "layered" phase where the face-centered-cubic (fcc) lattice
has
alternating Cu (spin up) and Au (spin down) planes stacked along the
(100) direction and there is a tetragonal distortion (the $L1_0$
structure observed experimentally \cite{Pearson}) which shortens the
antiferromagnetic bonds and stretches the ferromagnetic
bonds \cite{prl,Xi}.  This observation implies that the $CuAu$
transition
is in the same class as the triangular lattice Ising antiferromagnet
coupled to uniform lattice distortions.  Part of the physics entering
the EMT Hamiltonian for Cu-Au can therefore be understood
simply in terms of Ising
antiferromagnets on frustrated but deformable lattices.  The advantage
of EMT is that the model is {\it derived} from a microscopic description
of a given alloy and all the alloy-specific parameters are known.
However, the mapping of the physics onto a simple model helps in
understanding the physics of ordering in these alloys.

The other ingredient in EMT which is different from the usual Ising
model description
of alloys is the size of an atom.  This size is defined in EMT as
that radius of a
sphere within which each atom is neutral.  The size is therefore a
configuration
dependent quantity which depends on screening in a metallic
environment.
It has been argued \cite{prl} that the modulated $CuAuII$
phase \cite{Pearson} is a consequence of such size effects and it will
be shown here that these size effects lead to Ising models with
competing
interactions similar to ANNNI models \cite{selke}.  The $CuAuII$
phase
has a one-dimensional modulation of the order parameter along a cubic
direction and the period of the modulation, which varies as a
function of composition, is ten lattice constants at the 50-50
composition.  There is an alternate
mechanism which could lead to modulated phases in metallic alloys,
and this is driven by Fermi-surface instabilities \cite{Gyorffy}.  EMT,
in
its present form, cannot describe subtle Fermi surface effects and may
fail to describe modulated phases in alloys where this is the dominant
effect.  The model of kinetics that will be presented in this paper can
encompass both classes of modulated-phase alloys.  The model will be
derived for size-effect alloys such as $CuAu$ and the differences and
similarities with Fermi-surface alloys will be discussed.

\section{Model of Kinetics}
Pattern formation and kinetics of phase transitions are traditionally
described by the time-dependent Ginzburg--Landau model
\cite{Mazenko}.  For a non-conserved scalar order parameter
$(\psi )$, this model Langevin equation can be written as,
\begin{equation}
\label{langevin}
\frac{\pa\psi}{\pa t} = -{\frac{\d \cf}{\d\psi}} + \eta ({\rvec},t)
\end{equation}
where $\cf [\psi ]$ is the Ginzburg--Landau free energy functional, and
$\eta (r,t)$ is a Gaussian noise term reflecting fluctuations of the heat
bath
\cite{Gunton,Mazenko}:
\begin{eqnarray}
\label{noise}
<\eta({\rvec},t)> & = & 0,\nonumber\\
<\eta({\rvec},t) \eta({\rvec}^{\prime}, t^{\prime})>  & = &  k_B
T\delta({\rvec} - {\rvec}^{\prime} ) \delta (t - t^{\prime}) ~ .
\end{eqnarray}
It should be noted that $\cf [\psi ]$ also acts as a
Lyapunov functional; $\frac{\pa {\cf [\psi ({\rvec},t )]}}{\pa t}  \leq 0$.

Most
descriptions of kinetics in alloys have been based on the traditional
$\psi^4$ form of $\cf [\psi ]$.
In contrast, our discussion of kinetics will be based upon a $\cf [\psi ]$
{\it derived} from the microscopic EMT Hamiltonian.

We use as our prototype alloy the $CuAu$ (50--50) alloy which
exhibits modulated phases and, in equilibrium, has a first-order phase
transition accompanied by a lattice distortion.  The latter shows the
relevance of the coupling to phonons and uniform lattice distortions and
raises questions about the description of metastable and unstable states
and the concept of a spinoidal.  The presence of two first order
transitions, from the disordered to the modulated $CuAuII$ phase,
followed by the transition from $CuAuII$ to the zero temperature
unmodulated $CuAuI$ structure \cite{prl,Xi}, leads to unusual
quench-temperature
dependence of the kinetics and a
Swift-Hohenberg-like \cite{swift} description of the
early-stage kinetics.

The free-energy functional $\cf [\psi ]$ for the $CuAu$ alloy has been
constructed by numerically calculating the mean-field free energy from
the EMT Hamiltonian \cite{prl} and we will only briefly discuss the
procedure and the results in this paper.
The Hamiltonian includes uniform lattice distortions and the Ising
degrees of freedom, $H ({\{ e_{\alpha} \}}, {\{ s_i
\}})$.  Where $s_i$ are the Ising variable on lattice sites $i$, and
the $\{ e_{\alpha} \}$ are the set of uniform lattice
distortions.   The partition function for this model can be written as
\begin{equation}
\label{partit}
Z = \int {\Pi}_{\alpha} de_{\alpha} \int {\cal D}\psi \exp (-F[\psi
,e_{\alpha}])
\end{equation}
{\it ie}, one integrates over the lattice distortions and the order
parameter field $\psi$.  The Ginzburg-Landau functional $\cf [\psi]$
is
obtained by making a saddle-point approximation to the $e_{\alpha}$
integrals, {\it ie} minimizing $F$ with respect to these variables.  In
our numerical procedure, the mean-field free energy corresponding to
a
given set of $\psi$ and a given set of $e_{\alpha}$ was calculated from
the EMT hamiltonian and this function was then minimized with
respect to
the $e_{\alpha}$.  This is equivalent to making the
magnetothermomechanics \cite{schultz} approximation.  The
minimization of
$\cf [\psi]$ with respect to $\psi$ leads to the mean-field description
of the $CuAu$ ordering transitions \cite{prl}.  In this paper, this
Ginzburg-Landau functional is used to describe the kinetics of phase
transitions in these alloys.

The form of  $\cf [\psi ]$ for $CuAu$,
is most conveniently written
in momentum space \cite{footnot2}:
\begin{eqnarray}
\label{lg}
\cf [\psi ] & = & \sum_{\qvec} [a(T-T_0 ) + \o (\qvec) ]
\psi_{\qvec} \, \psi_{-{\qvec}} \nonumber \\
& + & \frac{ u^{\prime}}{4} \; \sum_{\qvec} \; \psi_{{\qvec}_1} \ldots
\psi_{{\qvec}_4} \; \d ({\qvec}_1 + {\qvec}_2 + {\qvec}_3 +{\qvec}_4 )
\nonumber \\
& - & \frac{u}{4} \; \left(\sum_{\qvec} \; \psi_{\qvec} \, \psi_{-{\qvec}}
\right)^2
\nonumber \\
& + & \frac{v}{6} \left(\sum_
{\qvec} \; \psi_{{\qvec}_1} \, \psi_{{\qvec}_2} \ldots
\psi_{{\qvec}_6} \; \d ({\qvec}_1 + \ldots {\qvec}_6 ) \right) \; .
\end{eqnarray}
where
$$
\o (\qvec) = e\: q^2_z + (q^2_\perp - q^2_0 )^2 -
q^4_0 \; .
$$
Here $q_{\perp}$ is the magnitude of a wavevector in the plane
perpendicular to the $CuAu$
ordering
which is taken to be in the $z$-direction.

Before
discussing the model of kinetics that follows from this Ginzburg-Landau
functional, we briefly discuss the applicability of this model to
equilibrium phase transitions in $CuAu$.  Because
of the form of $\o
(\qvec)$,
the Ginzburg-Landau functional, $\cf
[\psi ]$, is minimized by a one-dimensional periodic function, which
describes a configuration comprised of a periodic array of antiphase
boundaries separating regions with $\psi = \pm 1$.   This is the source
of the $CuAuII$ phase.   The appearance of stationary one-dimensional,
periodic patterns is also a feature of the Lyapunov functional
associated with the Swift-Hohenberg equation \cite{swift}.
The isotropic form of $\o (\qvec)$ implies that the one-dimensional
pattern is
rotationally invariant in both models.
However, the $CuAuII$ structure observed in
Monte Carlo simulations \cite{prl,Xi}, based on EMT, and in
experiments, has the
modulation locked along one of the cubic axes, and there are lattice
distortions accompanying this transition.
The unusual fourth
order term (3$^{\rm rd}$ term in Eq. (\ref{lg})) arises from the coupling
to
uniform lattice distortions. This term is responsible for the $CuAuI$
type
ordering
\cite{prl,Xi} which, as discussed earlier, is in the
same class as the Ising antiferromagnet
on an
elastic triangular lattice \cite{Kardar}.  This type of ordering
involves two broken symmetries, one corresponding
to the twofold
Ising degeneracy and one corresponding to the three possible strain
directions on the lattice.  The $CuAuII$ ordering involves a
lattice distortion  along the direction of the one-dimensional
modulation,  and an additional symmetry breaking in the choice
of this direction.   It is expected that the coupling to the lattice is
the source of the commensurate $CuAuII$ ordering.
In the Landau theory, presented earlier \cite{prl}, the modulation was
assumed to be along one of the cubic directions.  The effects leading to
the choice of this direction, within mean-field theory,  are currently
being investigated.  It is clear from our Monte Carlo simulations that
the EMT hamiltonian can describe the lattice distortions accompanying
the modulation \cite{prl,Xi}.

The function $\o(\qvec)$
can be measured in diffuse scattering
experiments \cite{hashimoto,ogawa}.
The EMT model would predict an isotropic distribution in the plane
perpendicular to the ordering direction ({\it cf} form of
$\o(\qvec)$ in Eq. (\ref{lg})) and this isotropic distribution
would have a peak displaced from the $CuAuI$ superlattice Bragg peak
by a
distance $q_0$.  Experimental measurements on $CuAu$ show an
essentially
isotropic distribution with a peak at $q_0 \simeq 2\pi /10 a$ where $a$
is
the lattice constant \cite{hashimoto}.   The EMT estimate for $q_0$ is
$\simeq 2\pi /4 a$ \cite{prl}.  This was, however, a very crude
numerical
estimate which was only meant to show that $q_0$ is nonzero.  The
essentially isotropic distribution of $CuAu$ is in sharp contrast with
diffuse scattering intensities in $Cu_3 Pd$ \cite{ogawa} where
the distribution has well defined peaks along the $q_x$ and $q_y$
directions.  These peaks have been explained on the basis of a
Fermi-surface mechanism \cite{stocks}.  The essentially isotropic
distribution in $CuAu$ suggests that a different mechanism is
responsible for the modulated phase.
The small anisotropy seen in
experimental diffuse scattering measurements in $CuAu$ would be
difficult to obtain from our numerical calculations based on EMT.  It
could also be that there is a Fermi surface contribution which is
missing from EMT that makes the distribution anisotropic.  Because of
the small anisotropy, it is reasonable to investigate the isotropic
model and discuss the consequences of the anisotropy. The isotropic
model is interesting in its own right since it implies an
the breaking of a continuous symmetry at
the $CuAuII$ transition, where the system chooses a direction of
modulation.

The above discussion shows that $\cf [\psi]$ is a good model for
describing kinetics of these alloys.
The
kinetic equation derived from the model  $\cf [\psi ]$, is
\begin{equation}
\label{kinetic}
\frac{\pa \psi ({\rvec},t)}{\pa t} = -
\left\{ \left[ a(T-T_0) - q^4_0 \right] - \frac{\pa^2}{\pa z^2} +
(q^2_0 + \nabla^2_\perp )^2 \right\} \psi ({\rvec},t) -
\frac{\d f}{\d \psi}  + \eta ({\rvec},t) ,
\end{equation}
where
$$
f [\psi ] \; = \; \frac{u^{\prime}}{4} \: \int \:
\psi^4({\rvec}) d{\rvec} - \frac{u}{4} \;
\left( \int \psi^2 ({\rvec}) d{\rvec} \right)^2 + \frac{v}{6} \: \int \:
\psi^6 ({\rvec})d{\rvec} \; .
$$

\section{Early Stage Kinetics}

The early stage kinetics of $CuAu$, following a quench,
is defined by a linear stability analysis \cite{cross}
of the
disordered, $\psi = 0$ state.  This implies neglecting the
$\d f/\d \psi$ term in Eq. (\ref{kinetic}).  The linear theory
is very different from the usual
model of a non-conserved order parameter \cite{cross,klein}.
The difference is evident from an analysis of the
solution to the linearized equations without the noise term.
The easiest route to this solution is through transformation to
momentum
space where the different modes decouple and the
solutions for $\psi(\qvec)$ are,
\begin{equation}
\label{cahn}
\psi(\qvec,t) = \exp (D(\qvec) t)
\end{equation}
where $ D(\qvec) = -a(T-T_0 ) -\o(\qvec)$.  This solution
shows that modes with $D(\qvec)$ greater than zero
grow exponentially with time.  Eq. (\ref{cahn})
is the generalization of the Cahn-Hilliard solutions describing
continuous ordering \cite{klein} to alloys with modulated phases.
If none of the eigenvalues, $D(\qvec)$, are greater than zero then there
is no continuous ordering and the disordered state is either stable or
metastable.  In
contrast to the Cahn-Hilliard solutions, the fastest growing mode  has
a
finite wavevector, $q_0$ and the growth rate depends only on the
magnitude of the wavevector and not its direction.  This is reminiscent
of Cahn-Hilliard description of spinodal decomposition but there the
wavevector of maximum growth depends on the
temperature \cite{Gunton,klein} whereas in our model, the linear
dynamics chooses a unique length scale which is a characteristic of the
system.
Since the interesting pattern
formations take place in the plane perpendicular to $q_z$, we will
analyze the equations in the $q_z = 0$ plane.  This
linearized $CuAu$ equation is then
identical to the linearized version of the two-dimensional, stochastic
Swift-Hohenberg
equation \cite{Elder};
\begin{equation}
\label{hohenb}
\frac{\pa \psi ({\rvec},t)}{\pa t} =
\left\{ \epsilon  -
(1 + \nabla^2_\perp )^2 \right\} \psi ({\rvec},t) ,
\end{equation}
if the control parameter $\epsilon$
is
identified with
${q_0}^4 - a(T-T_0 )$ and the natural wavevector $q_0$ is set to 1.
The Swift-Hohenberg equation is used to model
pattern-formation in
Rayleigh-Benard convection and has been studied extensively both in
one and two
dimensions \cite{Elder}.

The linear dispersion relation ($D(q)$) is
plotted in Fig. (1) for various temperatures.  Three different
temperature regimes can be identified.  For $T \geq T_0 + {q_0}^4 /a
$,
all modes decay and there is no continuous ordering, this is the regime
where the disordered state is metastable and ordering takes place by
nucleation.  For $(T_0 + {q_0}^4) < T < T_0 $, a band of wavevectors
centered around $q_0$ grows and at $T= T_0$, the $q=0$ mode
becomes
unstable. The $q=0$ structure corresponds to the unmodulated
$CuAuI$
phase.  In the Cahn-Hilliard description of spinodal decomposition, the
growth rate at $q = 0$ is always zero, and in its description of
continuous ordering, the $q=0$ mode always has the fastest growth
rate.
It should be remembered that the linear dispersion relation
is isotropic in the plane perpendicular to $q_z$ and all modes with a
given magnitude of $q$ grow at the same rate.  This is expected to give
rise to an
interesting morphology of the early stage ordering
process \cite{kleinfrac}.

Experimental sudies of kinetics are based on the
structure factor.  The structure factor obtained from the linearized
equation at a temperature within the second temperature regime is
shown
in Fig. (2a).  The plot shows the evolution of the structure
factor as a
function of time.  The peak of the structure factor remains stationary
as as a small band of wavevectors centered around this peak grows as
a
function of time.  The modes outside this band are seen to decay with
the growth of order.
At a lower temperature, the structure factor at $q = 0$ would start to
grow, but the growth rate predicted by the linear theory always has its
maximum at $q = q_0$.
The inclusion of the Gaussian noise term changes the
form of the structure factor (generalization of the
Cahn-Hilliard-Cook theory \cite{klein}) and can be written as
\begin{equation}
\label{struc}
S(\qvec,t) = S_0 (q) \exp (2D(\qvec)t) + \frac{k_B T}{2D(\qvec)}
(\exp (2D(\qvec)t) - 1)
\end{equation}
where $S_0 (q)$ is the structure factor in equilibrium at the
temperature prior to the quench.   The  inclusion of the Cook term
introduces minor modifications of the structure factor which are
illustrated in Fig. (2b).
These plots show that an
experimental
investigation of the early stage kinetics of $CuAu$ in this temperature
regime should show a peak growing at a finite wavevector (measured
with
respect to the superlattice Bragg peak) and there should be a ring of
wavevectors at which the structure factor is a maximum.  The question
of
whether the early-stage description is ever valid in these systems can
only be  answered after a detailed study of the complete non-linear
equations or from Monte Carlo simulations of the appropriate model
Hamiltonian.  It has been argued that the length of time over which
the early stage kinetics
remains valid  is given approximately by $1/D(q_0)$ and depends
logarithmically on the range of interaction \cite{binder}.  In our
model, we have an effective infinite-range four-spin interaction arising
from the coupling to the lattice distortions and this might imply that
the range of validity of early-stage kinetics will be significantly
enhanced in these systems.  The
evolution of the lattice distortion as a function of time is a novel
feature of these systems and experimental investigations should shed
light on the coupling between the evolution of order and the evolution
of the lattice distortions.

\section{Discussion of Nonlinear Effects}

The effect of nonlinearities may change the kinetics in various ways.
In the
simplest case, the nonlinearities mainly come in to dampen the
exponential
growth.   Some simple considerations of the nonlinearities in our model
indicate
that there can be more significant differences between the linear and
nonlinear
dynamics.
The periodic solutions to which the $\psi = 0$ state becomes linearly
unstable have wavevectors lying in the range $[k_1 , k_2]$, where
\begin{equation}
\label{band}
k_{1,2} = ({q_0}^2 \pm \sqrt {({q_0}^4 - a(T-T_0 ) )} )^{1/2}
\end{equation}
For temperatures close to but less than $T_1 = T_0 + {q_0}^4/a $, the
width
of the
band of
wavevectors grows as  $|T - T_1 |^{1/2}$.     As the temperature
approaches $T_0$, the $q=0$ mode becomes unstable and below $T_0$,
the
width grows as  $(T-T_0 )$ for $T$ close to $T_0$.
This band of wavevectors, chosen by the linear dynamics, usually define
the range in which the stationary periodic solution is found at a given
temperature; {\it ie}, one could examine the periodic solutions lying in
the range $[k_2 , k_1 ]$, and the solution with the lowest free energy
becomes the selected pattern \cite{goldenfeld}.
However, because of the negative fourth
order term in $\cf [\psi]$, there is a first order transition from the
disordered to the ordered periodic phase.  This implies that, in a
given temperature range, periodic solutions which are outside the band
predicted by the linear dynamics could arise through a nucleated
process
and could become the stationary solutions.  The condition that a order
parameter of the form $\psi({\rvec})= \psi \exp (i{\qvec}\cdot{\rvec})
$
is a solution to $$
{\frac{\d \cf}{\d\psi}} = 0$$ is $$D
(\qvec) \geq
-\frac{(u-u^{\prime})^2}{8v}$$ compared to the condition for linear
stability, $D(\qvec) \geq 0$.
For example, the $q=0$ state becomes the stable stationary solution at
a
temperature higher than its instability temperature $T_0$.  The
role of nonlinearities is different in this model than in the
Swift-Hohenberg model.   Since the negative fourth order term is a
consequence of the coupling to the lattice \cite{prl,Kardar}, the
investigation of non-linear dynamics has to take proper account of this
coupling.  The other crucial difference between this model and the
Swift-Hohenberg model, is that the underlying lattice in
$CuAu$ introduces a second length scale comparable to the
length scale chosen by the linear dynamics ($2\pi /{q_0}$) and may be
instrumental in locking in the direction of modulation.  These effects
are , in
principle, all there in the
nonlinear terms of our model equations if we are careful in keeping the
Umklapp
terms involving nonzero reciprocal lattice vectors in Eq. (\ref{lg}).
However,
it might be more profitable to investigate these effects directly
in an effective lattice model.

Numerical solutions of the full non-linear Swift-Hohenberg equations
have shown that the lamellar patterns exhibit phases which are akin
to nematic
and smectic ordering in liquid crystals \cite{Elder}.  In $CuAu$ alloys,
the
lamellar patterns have a length scale comparable to the underlying
lattice and
an analogy maybe drawn between these systems and nematic ordering
in the presence
of an underlying lattice \cite{Selinger}.   The late stage growth laws in
Swift-Hohenberg systems is expected to be different from usual models
of
non-conserved order parameters since the wavevector defining the
modulation has a
continuous symmetry \cite{Elder}.  It would be interesting to study the
late stage
growth laws in our model which has the extra feature of coupling to the
lattice.

\section{Lattice Hamiltonian}

We have argued that the study of non-linear dynamics is $CuAu$-like
alloys may be best approached through a discrete lattice Hamiltonian.
With this in mind, a lattice Hamiltonian has been constructed which
has
the salient features of the EMT Hamiltonian when applied to
ordering on a deformable lattice \cite{footnot3}.  The model is an Ising
antiferromagnet on an elastic lattice with an unusual pair interaction
term which arises from the size difference between the two chemical
species making up the alloy.  The source of this term in EMT is the
density dependent cohesive energy \cite{emt,Xi}.  When expanded in
terms of
the spin variables, a part of this energy gives rise to a term in
the Hamiltonian
which is an antiferromagnetic interaction between two atoms which
share
a common nearest neighbor \cite{future}.  This
interaction depends on the bond lengths and therefore can be different
along different lattice directions in the presence of deformation.  The
strength of the antiferromagnetic interaction depends on the
size-difference of the two atoms and is zero for atoms of the same size.
This model can be written as,
\begin{equation}
\label{hamil}
H  =  {\sum}_{\alpha} J(1-\epsilon e_{\alpha}) {\sum}_{<ij>_{\alpha}}
s_i s_j
+ {\sum}_{\alpha \beta} M_{\alpha \beta} e_{\alpha} e_{\beta}
 +  {\sum}_{\alpha} K \ {\sum}_i ( {\sum}_{<ij>_{\alpha}} s_j )^2 ~ .
\end{equation}

The parameters can all be obtained from EMT, given a specific alloy,
however, we will treat this as a general model with arbitrary
parameters.  This should then form a generic, minimal model for alloys
where Fermi surface effects do not dominate.  The first
two terms in Eq. (\ref{hamil}) describe an Ising model coupled to
uniform strain through the coupling constant $\epsilon$ and an elastic
Hamiltonian defined by the tensor $M_{\alpha \beta}$.  This model for
$J
> 0$ has been analyzed on the triangular lattice \cite{Kardar} and gives
rise to a first order transition from a disordered to a striped phase.
The last term in the Hamiltonian is the unusual term arising from size
effects.  Since $K \geq 0$, this term favors a spin configuration where,
for each site, the sum of the nearest neighbor spins along a given
direction add up to zero.  In terms of atoms, this says that when atoms
of different sizes are present, it is favorable to have a big atom and a
small atom as nearest neighbors.   Neither an antiferromagnetic, nor
a
striped phase meets the optimum demand of the $K$ term and a
mean-field
argument shows
that, on a triangular lattice, modulated phases occur beyond a
critical value of $K$ \cite{future}.  This is reminiscent of the ANNNI
model.  The one crucial feature which distinguishes the current model
from
the ANNNI model is that all order-disorder transitions are first-order
because of the coupling to the lattice and makes it a more attractive
model for describing alloys.
Qualitatively, the source of the
modulation can be inferred from looking at the striped phase of the
triangular lattice antiferromagnet
which has alternating chains of up
and down spins and the sum over nearest neighbor spins for each
direction adds up to two.  If a modulation is introduced along the
ferromagnetic direction, such that after N spins, the spin up and spin
down chains are interchanged, then the spins at the domain boundaries
have nearest neighbor configurations which do satisfy the $K$ term
requirement.  The reason for the choice of the ferromagnetic direction
has its origin in the preferred lattice distortion.  Therefore in spite
of the $K$ term being isotropic, a unidirectional modulation occurs.
This scenario is very similar to the $CuAuII$ ordering and
investigations on the fcc lattice are currently underway.  At this
stage, we would like to present this model as an interesting variation
on competing interactions models and suggest that the study of kinetics
of this model would lead to novel features which can be tested
through experimental studies of alloys.

\section{Conclusion}

As mentioned at the beginning of the paper, there is a class of
modulated alloys
where the modulation is due to Fermi-surface effects \cite{stocks}.   In
these
alloys, the $q-$dependence of the second order term in $\cf [\psi]$ is
not
isotropic; {\it ie}, $\o (\qvec)$ has well defined minima along directions
chosen
by the Fermi-surface nesting vectors.  The linear dynamics in these
alloys
chooses a length scale and a set of directions and the symmetry
breaking that
takes place is not continuous but discrete, Ising like.  Also, the role of
the
lattice distortions are expected to be minimal and therefore the
nonlinearities
may play a simpler role and the late stage kinetics may follow the same
scaling
laws as in the usual models.


A complete description of ordering kinetics in alloys has to take into
account the effect of phonons and local lattice distortions.
Allowing for fluctuations in bond lengths also leads to an
additional size-effect derived term in the lattice Hamiltonian.
Under certain circumstances, this can give rise to a long-range pair
interaction\cite{onuki}.  The EMT prediction for this strain-mediated
pair interaction term and its effects on kinetics is currently being
investigated.

In conclusion, we have presented a model for describing kinetics of
ordering in
alloys which have modulated phases.  The linear theory shows that the
kinetics is
similar to that of Swift-Hohenberg models, but the nonlinear terms are
distinctly
different from the Swift-Hohenberg equations.  A lattice Hamiltonian
which
captures the essential features of the microscopic model of these alloys
has been
constructed and simulations based on this model should be helpful in
understanding both the statics and dynamics of ordering in these alloys.

\section{Acknowledgements}

The author wishes to thank W. Klein, Karl Ludwig, Duane Johnson,
Frank Pinski and Mohan Phani for many helpful conversations. This work
was supported
in part by the NSF grant DMR-9208084.

\begin{figure}
\caption{The linear dispersion relation $D({\bf q})$
vs. $|q_\perp |$.
The dispersion is isotropic in the $|{\bf q}_\perp |$
plane.  Four
temperatures are shown with the bottom most curve at the temperature
$T_1 = T_0 +{q_0}^4 /a$ and the topmost curve at the temperature
$T_0$.}
\label{fig1}
\end{figure}

\begin{figure}
\caption{(a) The evolution in time of the structure factor
obtained from the linear theory without
the noise term.  The plots show $S({\bf q})$
at a temperature $T$ such
that $T_1 \leq T \leq T_0$.  The crosses are at the earliest time and
the squares at the latest time.
(b) The evolution in time of the structure factor
predicted by the linear theory with the noise term included.  The
parameters used are those appropriate for CuAu.}
\label{fig2}
\end{figure}


\begin{references}
\bibitem{footnot2}  In writing this form it has been assumed that the
wavevectors
are much smaller that a reciprocal lattice vector, if this condition is not
satisfied than the constraints imposed by the delta functions have to be
replace
by constraints where the $q$-vectors add up to a reciprocal lattice
vector.
These are the so-called Umklapp terms.
\bibitem{footnot3}  The EMT Hamiltonian can describe alloy phases
which are not
on a lattice, here we look at a restricted Hamiltonian which describes
alloy on a
deformable lattice.
\end{references}
\end{document}